# Transport and superconducting properties of Fe-based superconductors: SmFeAs(O$_{1-x}$F$_x$) versus Fe$_{1+y}$(Te$_{1-x}$,Se$_x$)

M Tropeano[a,d], I Pallecchi[a], M R Cimberle[b], C Ferdeghini[a], G Lamura[a], M Vignolo[a], A Martinelli[a], A Palenzona[a,c] and M Putti[a,d],

[a] CNR/INFM-LAMIA Corso Perrone 24, 16152 Genova, Italy
[b] CNR-IMEM, Dipartimento di Fisica, Via Dodecaneso 33, 16146 Genova, Italy
[c] Dipartimento di Chimica e Chimica Industriale, Università di Genova, Via Dodecaneso 31, 16146 Genova, Italy
[d] Dipartimento di Fisica, Università di Genova, Via Dodecaneso 33, 16146 Genova, Italy

**Abstract**

In this work we carry out a direct comparison between transport and superconducting properties - namely resistivity, magnetoresistivity, Hall effect, Seebeck effect, thermal conductivity, upper critical field - of two different families of Fe-based superconductors, which can be viewed in many respects as end members: SmFeAs(O$_{1-x}$F$_x$) with the largest T$_c$ and the largest anisotropy and Fe$_{1+y}$(Te$_{1-x}$,Se$_x$), with the largest H$_{c2}$, the lowest T$_c$ and the lowest anisotropy. In the case of the SmFeAs(O$_{1-x}$F$_x$) series, we find that a single band description allows to extract an approximated estimation of band parameters such as carrier density and mobility from experimental data, although the behaviour of Seebeck effect as a function of doping demonstrates that a multiband description would be more appropriate. On the contrary, experimental data of the Fe$_{1+y}$(Te$_{1-x}$,Se$_x$) series exhibit a strongly compensated behaviour, which can be described only within a multiband model.

In the Fe$_{1+y}$(Te$_{1-x}$,Se$_x$) series, the role of the excess Fe, tuned by Se stoichiometry, is found to be twofold: on one hand it dopes electrons in the system and on the other hand it introduces localized magnetic moments, responsible for Kondo like scattering and likely pair-breaking of Cooper pairs. Hence, excess Fe plays a crucial role also in determining superconducting properties such as the T$_c$ and the upper critical field B$_{c2}$. The huge B$_{c2}$ values of the Fe$_{1+y}$(Te$_{1-x}$,Se$_x$) samples are described by a dirty limit law, opposed to the clean limit behaviour of the SmFeAs(O$_{1-x}$F$_x$) samples. Hence, magnetic scattering by excess Fe seems to drive the system in the dirty regime, but its detrimental pairbreaking role seems not to be as severe as predicted by theory. This issue has yet to be clarified, addressing the more fundamental issue of the interplay between magnetism and superconductivity.

## 1. Introduction

Soon after the discovery of superconductivity in Fe-based compounds [1], they have been indicated as the new unconventional superconductors which could compete with high-T$_c$ cuprates for application purposes. In Fe-based materials, several common features have been identified: in general, superconductivity appears upon chemical doping of an antiferromagnetic parent compound, with an optimal doping level which yields the largest transition temperature; the parent compound is an almost compensated semimetal, whose Fermi surface fulfills the nesting condition; the presence of Fe, which used to be considered detrimental for superconductivity, may be a crucial ingredient therein, instead; spin-fluctuation mediated pairing has been suggested [2]; the crystal structure is layered and usually composed of functional blocks playing the roles of charge reservoirs and high mobility planes of Fe square lattices. Yet, as the research on these new compounds has developed, it has appeared more and more clear that the physical mechanisms into play form a more complex scenario and that among different Fe-based phases there are significant, maybe crucial, differences.

This suggests the idea that an identical theoretical framework cannot be used to account for superconducting mechanisms in all Fe-based superconductors. In particular, the so called "1111" phase with general chemical formula REFeAsO (RE=La, Sm, Nd, Ce...) is the one with the largest $T_c$ of 56K [3], while the so called "11" phase of iron chalcogenides (FeCh, Ch=chalcogenide) is appealing due to its simple structure, the possibility of growing fairly large single crystals [4] and high quality epitaxial films [5] and the reduced toxicity of its constituents compared to As. The "11" phase presents several peculiarities as compared to the "1111" phase, namely the non-collinear orientation of anti-ferromagnetic ordering vector and nesting vector [6,7] and no clear signatures in favor of a spin-density-wave (SDW) gap [8,9,10]. Besides these possibly fundamental features which may be very important clues of a different pairing mechanism, there are other differences between "11" and "1111" phases relevant for applications. The dependence of $T_c$ on chemical substitution is less steep in Fe(Te,Se), with possible advantages regarding possible phase separation at grain boundaries for weak link behaviour of critical current. Upper critical fields, $B_{c2}$, of the two families are extremely high, and present different behaviours [11]. In the "1111" phase, $B_{c2}$ is strongly anisotropic and for H parallel to the c-axis it exhibits an upward curvature with decreasing temperature, reminiscent of two-band behaviour in $MgB_2$.[12] In the "11" phase, the anisotropy is rather low and it is rapidly suppressed with decreasing temperature. This is due to an anomalous downward curvature, more evident for H parallel to the ab-plane, which is likely related to the Pauli paramagnetic limit [11]. The coherence lengths in both the families are rather small, if compared with the interlayer distance, which is the largest and the smallest in the "1111" and "11" phases, respectively. Indeed, the analysis of thermal fluctuations, not negligible in both the cases, allows to get information of the dimensional character of superconductivity: 2D in the "1111" phase [13] and 3D in the "11" phase [11].

In this work we present a one to one comparison between electrical and thermal transport properties of the "1111" and "11" families, which can be considered as end members of the Fe-based superconductors. In particular, we show resistivity, magnetoresistivity, Hall effect, Seebeck effect and thermal conductivity data for both phases. We discuss the relevant behaviors in order to extract information on parameters like mobility and carrier density that are important to describe the normal and superconducting state properties. Finally we single out the most significant differences mainly related to the strongly compensated nature of "11" in comparison with "1111". Superconducting properties are finally discussed and the important role of magnetic moments induced by excess Fe in the "11" phase is pointed out.

**2. Experimental details**

SmFeAs($O_{1-x}F_x$) (Sm-1111) polycrystalline samples were synthesized as reported in ref. [14] and their magnetic [15], transport [16] and thermal [17] properties were thoroughly investigated. We present here transport and thermal properties of three of them: undoped (x=0), underdoped (x=0.075) and optimally doped (x=0.15) samples.

Samples of $Fe_{1+y}(Te_{1-x},Se_x)$ (Fe-11) with $x$ = 0.00, 0.025, 0.05, 0.075, 0.10, 0.15, 0.20 and x=0.50 were prepared by means of solid state reaction among stoichiometric amounts of pure elements (Fe 99.99+%, Se 99.9% and Te 99.999%) with a two step procedure. First a mixture of the starting elements is reacted in a Pyrex tube at 400–450 °C for 1–2 days, then these first products are ground, pelletized and heated at 800 °C in a evacuated $SiO_2$ tube for 7–8 days. All the operations are carried out in a glove box where $O_2$ and $H_2O$ were less than 1 ppm.

On the Fe-11 samples, neutron powder diffraction (NPD) analysis was carried out at the Institute Laue Langevin (Grenoble – France) and the results of thermo-diffractograms and high resolution NPD patterns will be discussed and published elsewhere together with complete Rietveld refinement data [18].

Transport properties measurements were carried out with a Physical Properties Measurement System (PPMS, Quantum Design) in the temperature range 5-300K and in field up to 9T. Hall coefficients ($R_H$) were determined measuring the transverse resistivity at selected temperatures sweeping the field from -9T to 9T. Seebeck coefficient (S) and Thermal Conductivity ($\kappa$) were measured with the PPMS Thermal Transport Option with a 0.2 K/min heating rate.

For the purposes of discussing the transport properties in this compounds, cell parameters and amount of excess Fe are reported in Table 1. Due to the substitution of Te with smaller Se the cell volume progressively decreases and the percentage of reduction of the cell parameter $c$ is higher than that of $a$. The occupancies at the Fe sites were refined and it is evident that occupation of Fe decreases with Se substitution as recently reported in ref. [19]. Following this trend we assume for x=0.5 sample the least amount of excess Fe, since for NPD analysis was not carried out in that case.

| x | Excess Fe y | $a$ (Å) | $c$ (Å) | $T_{SM}$ (K) | $T_c$ (K) | $dB_{c2}/dT$ (T/K) |
|---|---|---|---|---|---|---|
| 0.00 | 0.05 | 3.8219(1) | 6.2851(1) | 72.7 | - | - |
| 0.05 | 0.04 | 3.8184(1) | 6.2617(1) | 50.8 | 11.0 | - |
| 0.10 | 0.03 | 3.8160(1) | 6.2381(1) | - | 11.9 | -16.3 |
| 0.15 | 0.02 | 3.8133(1) | 6.2116(1) | - | 12.7 | - |
| 0.20 | 0.02 | 3.8114(1) | 6.1843(1) | - | 13.6 | -12.9 |
| 0.50* | - | 3.8041(2) | 6.0530(4) | - | 15.6 | -8.2 |

**Table 1**: structural parameters from Reitveld refinement analysis for selected composition in Fe-11 compounds. $T_{SM}$ is evaluated as the maximum of the resistivity slope; $T_c$ is evaluated at the 90% of the transition; $dB_{c2}/dT$ is evaluated by magnetoresistivity measurements up to 9T.
*Structural parameters of x=0.5 sample are evaluated by X-ray diffraction

## 3. Transport properties of Sm-1111

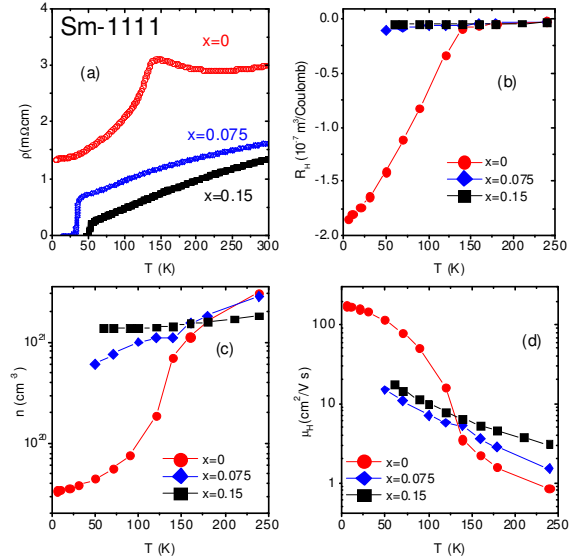

**Figure 1:** Resistivity and Hall effect measurements for Sm-1111 samples: (a) resistivity, (b) Hall resistance, (c) Carrier density and (d) Hall mobility versus Temperature

Figure 1(a) shows the resistivity of the selected samples: the undoped sample shows a drop of resistivity in correspondence of the SDW order whose transition temperature, $T_N$=131 K, is defined as the maximum of the resistivity derivative. No magnetic transition has been shown by x=0.075 and x=0.15 superconducting samples, whose superconducting critical temperatures ($T_c$) are 33 K and 51.5 K respectively.

In figure 1(b) the Hall coefficient, $R_H$, is plotted. In a two band framework $R_H$ is given by:

$$R_H = \frac{(\mu_h^2 n_h - \mu_e^2 n_e)}{(|\mu_h| n_h + |\mu_e| n_e)^2} \quad (1)$$

where $n_{e(h)}$ and $\mu_{e(h)}$ represent the carrier concentrations and mobilities of the electron(hole) bands. $R_H$ is negative in all the samples, indicating that electrons play the main role in the conduction of these compounds; assuming, as a first approximation, a predominant e-type conduction, it is possible to evaluate the electron density as $n = 1/e|R_H|$ where $e$ is the electron charge. This is plotted in fig. 1(c). Given the electron density donated by F, $\delta n = x/V_{cell}$ where $V_{cell}$ is the unit cell volume, the optimally doped sample should have an excess of charge with respect to the undoped of about $\delta n \approx 10^{21}$ cm$^{-3}$. Actually at 240 K we find the opposite: $n$ is around $3\times 10^{21}$ cm$^{-3}$ and $2\times 10^{21}$ cm$^{-3}$ for the x= 0 and x=0.15 samples, respectively. Such discrepancy, even if partially due to the uncertainty on $R_H$ data related to the polycrystalline nature of the samples, evidences the limits of the single-band approximation for precise quantitative values, nevertheless the temperature trend of $n$ is reliable. With decreasing temperature, while the carrier density of the optimally doped sample is rather constant, in the undoped sample $n$ is strongly suppressed below $T_N$, as an effect of the opening of the SDW gap. The Hall mobility evaluated as $\mu_H = R_H/\rho$ is plotted in figure 1(d). In a single-band approximation $\mu_H$ represents the mobility of the carriers, and the reported behaviours, looking rather reasonable, corroborate this view: with decreasing temperature $\mu_H$ increases rather smoothly in the doped samples, whereas in the undoped one it strongly increases below $T_N$. Such steep increase, that is yielded by the drop of resistivity below $T_N$ despite of the reduction of carriers, has been considered as an evidence of carrier correlation [16].

In order to confirm the outlined picture, magnetoresistivity (MR) measurements of the undoped sample are shown in figure 2. MR strongly depends on temperature: it is rather large (15%) at 5 K, but it rapidly decreases with increasing temperature and becomes negligible at $T_N$. In a single band system the temperature dependence of MR scales as the product of the magnetic field times the mobility as stated by Kohler rule. In the inset of figure 2, Δρ/ρ(0) is plotted as a function of $B\mu_H$. All the curves fairly overlap, which proves that $\mu_H$ is a good evaluation of the carrier mobility.

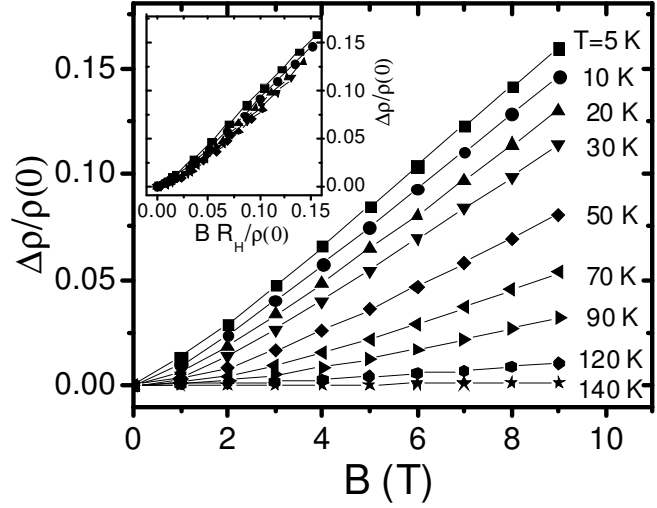

**Figure 2:** $\Delta\rho/\rho(0) = [\rho(B) - \rho(0)]/\rho(0)$, measured at fixed temperature with increasing B of SmFeAsO. Inset Kholer's plot. $\Delta\rho/\rho(0)$ plotted as a function of $BR_H/\rho(0) = B\mu_H$.

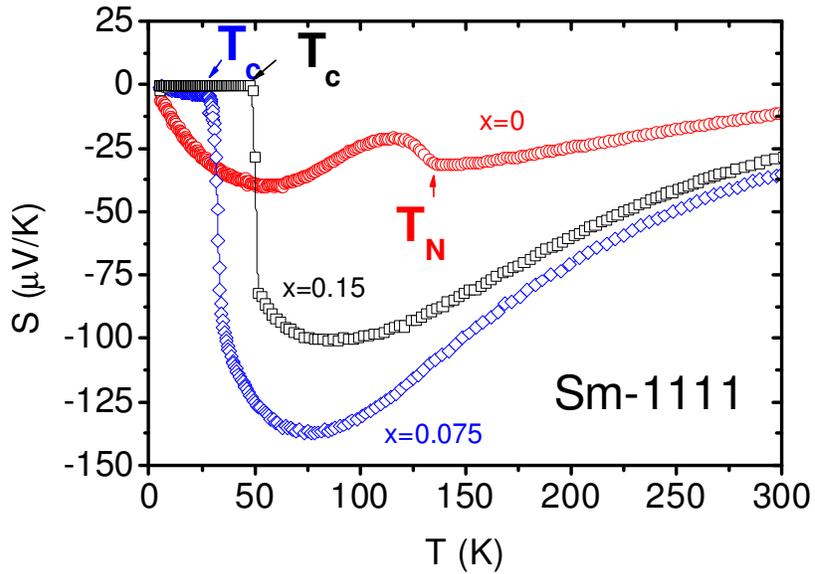

**Figure 3:** Seebeck coefficient of Sm-1111 samples versus T.

The Seebeck coefficient, *S*, of the three samples is shown in fig. 3. *S* is negative over the entire temperature range, in agreement with the Hall coefficient and data reported in ref. [20]. In the undoped sample *S* presents a decrease in absolute value below $T_N$ that can be understood within a free electron model where *S* is given by:

$$S(T) = -\frac{\pi^2}{3}\frac{k_B}{|e|}k_B T\left[\frac{N(0)}{n} + \frac{1}{\mu}\frac{d\mu}{dE_F}\right] \qquad (2)$$

where $E_F$ is the Fermi energy and $N(0)$ the density of states. The first term is always positive, while the second one, which can be qualitatively related with the variation of mobility with doping

($d\mu/dx \propto d\mu/dE_F$) changes sign with temperature. Indeed, $\mu_H$ of the x=0 and x=0.075 samples cross at $T_N$ (see fig. 1(d)), which means that $d\mu_H/dx > 0$ above $T_N$ and $d\mu_H/dx < 0$ below $T_N$. This explains nicely the anomaly below $T_N$ and again evidences that the single-band approximation is largely working in these compounds. However, looking at figure 3, a non monotonic behaviour of the Seebeck effect with doping can be emphasized: $|S|$ increases in absolute value from x=0 to x=0.075 and then decreases for x=0.15. This can be understood considering the contribution of two bands: in this case $S$ is expressed as:

$$S = \frac{\sigma_h|S_h| - \sigma_e|S_e|}{\sigma_h + \sigma_e} \qquad (3)$$

where $\sigma_{e(h)}$ and $S_{e(h)}$ are the contributions of electrons (holes) to the electrical conductivity and Seebeck coefficient, respectively. The x=0 sample shows smaller $S$ values since it is the most compensated, whereas electron doping strengthen the electron contribution and $|S|$ increases (x=0.075). With further electron doping (x=0.15), $|S|$, which is proportional to $1/n$ (see eq.(2)), decreases again.

In summary, by a close inspections of the Hall effect, magnetoresistivity and Seebeck effect data it comes out that single-band relationship can be reliably used for quantitative analysis despite the multiband nature of the Sm-1111.

## 4. Transport properties of Fe-11

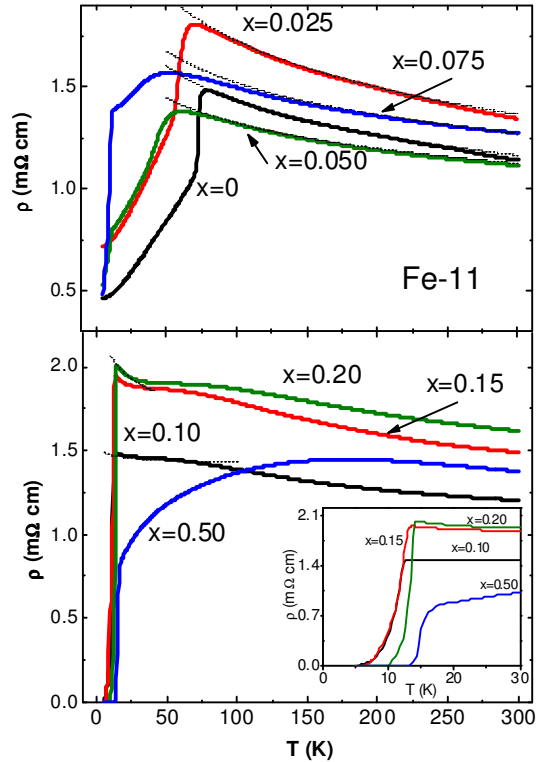

**Figure 4:** resistivity versus temperature curves of our Fe-11 samples. Low and high Se content samples are plotted in the upper and lower panel respectively. Dashed lines are logarithmic fit curves. The inset shows the transition region for the superconducting samples.

In figure 4, we present resistivity measurements of Fe-11 polycrystalline samples with x ranging from 0 to 0.5. In particular, in the upper and lower panels the series with low and high Se content, respectively, are displayed separately, for clarity sake. In all cases the normal state resistivity is of the order of 1 mΩcm, similar to the one of the Sm-1111 series. In the upper panel, the parent compound x=0 shows a negative temperature dependence above around 72K, where the structural and magnetic (SM) transitions occur simultaneously [7] At the transition, the resistivity undergoes and abrupt jump to a lower value and a exhibit a metallic behavior down to the lowest temperatures. This behavior is akin the one of the Sm-1111 parent compound, but for the abrupt jump, peculiar of Fe-11. With increasing Se substitution the transition is gradually smoothened and the transition temperature ($T_{SM}$) is lowered down to 26K in the x=0.075 sample. In this latter sample, the onset of the superconducting transition appears at 12K, even if resistivity does not vanish yet at 5K. In the lower panel, the samples with x from 0.1 to 0.5 present a superconducting transition with maximum $T_c$ of 15.6 K for the x=0.5 sample. All the transition temperatures are reported in Table I, where they are defined at 90% of the normal state resistivity. Also the structural transition temperatures, defined as the maximum of the resistivity derivative, are reported in Table I. It can be seen that the superconducting transition temperature increases very weakly and monotonically with the Se content, whereas the structural transition temperature decreases monotonically. In the Se content range x=0.050-0.075, magnetic order and superconductivity coexist in the same sample [21]. The x=0.1, 0.15 and 0.2 samples have a negative temperature dependence above the transition, pointing to a larger degree of localization as compared to the Sm-1111 phase, where metallic behavior is observed. Also Seebeck [22] and magnetic susceptibility [19] measurements on Fe-11 single crystals confirm this more localized character. Such temperature dependence is roughly described by a logarithmic law, as shown by dashed fit lines in figure 4. The x=0.5 sample exhibits a quasi metallic behavior.

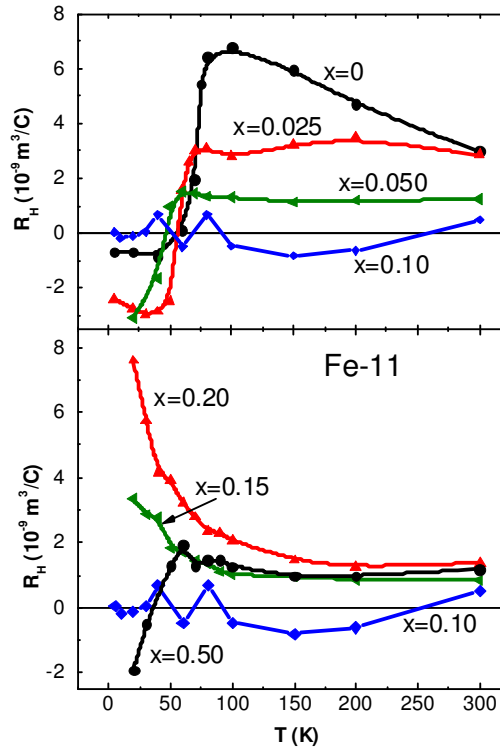

**Figure 5:** Hall resistance versus temperature curves of our series of Fe-11 samples; the low Se content samples are plotted in the upper panel and the higher Se content ones in the lower.

In figure 5, Hall resistance curves are plotted as a function of temperature for the whole series of samples. Again, the low and high Se content are displayed in two different panels for clarity. The low Se samples in the upper panel exhibit in general a weak temperature dependence, except for abrupt jumps at $T_{SM}$. In the parent compound x=0 $R_H$ is positive and increases with decreasing temperature down to $T_{SM}$; at the transition it undergoes an abrupt jump from $6·10^{-9}$ m³/C to negative value around $-0.7·10^{-9}$ m³/C. No evidence of carrier condensation below $T_{SM}$ can be drawn, oppositely to the case of the undoped Sm-1111 sample. The behavior of the x=0.025 and 0.050 samples is qualitatively very similar. On the contrary, the $R_H$ of the x=0.1 sample is much smaller and multiple crossings from positive to negative values are seen as a function of temperature. As for the higher Se content samples, $R_H$ is positive above ~50K with a value 1-2·10$^{-9}$ m$^3$/C; at lower temperature it starts increasing steeply for the x=0.15 and x=0.20 samples and decreasing to negative values for the x=0.50 samples. A similar variety of low temperature behaviors have been reported in Fe-11 epitaxial films [23].

The single band description applied to the Sm-1111 phase data is not justified in this case due to the almost compensated multiband character that makes the interpretation of $R_H$ curves harder. Clearly, the sign of $R_H$ points to the sign of the dominant carriers, however the balance is critically reversed by Se substitution and temperature. The low Se samples are dominated by holes at high temperature and by electrons at low temperature, with an abrupt crossover of the dominant band at the $T_{SM}$. The x=0.1 sample presents several changes of sign in $R_H$; in a two-band schematization, this indicates that the quantity $\left(\mu_h^2 n_h - \mu_e^2 n_e\right)$ is almost vanishing and liable to sign changes (see eq.(1)). For larger Se content the hole conduction prevails.

Since in this strongly compensate case $\mu_H$ is not expected to be related to the carrier mobility, magnetoresistivity has been investigated. In figure 6, we present normal state MR data as a function of temperature for the whole series of samples. In sharp contrast with the corresponding data of the Sm-1111 phase, both positive and negative magnetoresistivity is observed. MR curves exhibit B$^2$ dependence at all temperatures (not shown), both in the case of negative and positive dominant contributions and its magnitude is extremely small, namely below 0.3% at 9T. Below $T_{SM}$ the x=0 parent compound has positive magnetoresistivity, increasing with decreasing temperature. At the transition a sharp peak of magnetoresistance is measured (see inset of figure 6); this suggests a dependence of $T_{SM}$ on the field that should be confirmed by specific investigation. After a change of sign around ~100K it returns positive. For the other samples, the magnetoresistance is mainly negative and its magnitude decreases with increasing temperature. For the x=0.5 sample, the positive contribution is again dominant at high temperature.

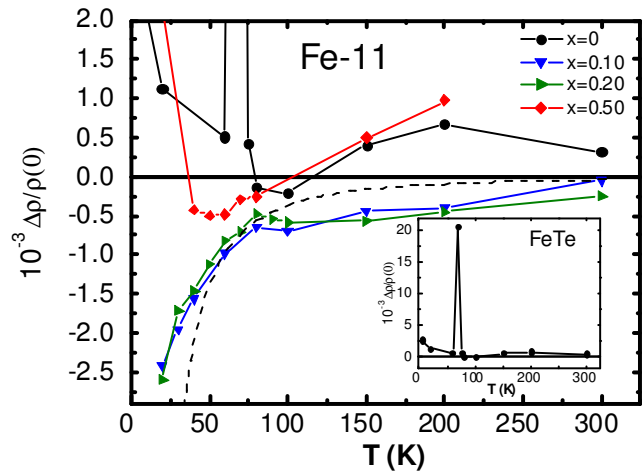

**Figure 6:** (ρ(B)-ρ(0))/ρ(0) as a function of temperature of our series of Fe-11 samples; the dashed line is the qualitative behaviour of Kondo contribution, calculated assuming a magnetic moment of localized impurities S=5/2. In the inset, the whole curve of the Fe$_{1+y}$Te sample is shown.

In a two band description, cyclotronic magnetoresistance is always positive and can be expressed as:

$$\frac{\rho(B)-\rho(0)}{\rho(0)} = \frac{\sigma_h \sigma_e (\mu_h - \mu_e)^2 B^2}{\sigma_h^2(1+\mu_e^2 B^2)+\sigma_e^2(1+\mu_h^2 B^2)+2\sigma_h\sigma_e(1+\mu_e\mu_h B^2)} \qquad (4)$$

Thus we tentatively attribute the positive magnetoresistance to the usual cyclotronic mechanism and the negative contribution to the Kondo mechanism, due to magnetic impurities scattering. The latter is justified because of the $B^2$ magnetoresistivity behaviour and the logarithmic temperature dependence of resistivity beyond the MS transitions. Also the temperature dependence of the magnetoresistivity confirms this hypothesis: dashed lines in figure 6 represent a typical behaviour of Kondo magnetoresistivity calculated using the formalism of ref. [24]. Excess Fe is indeed present in our samples (see table 1) and it is found to decrease monotonically with increasing Se. Density functional calculation [7] and neutron diffraction studies [25] established that excess Fe provides localized magnetic moments ~2.5$\mu_B$ strongly contributing to magnetic scattering.

In figure 7, we present Seebeck effect data of selected samples of the Fe-11 series. It is apparent that the overall behaviour is completely different from that of the Sm-1111 series. In the high temperature regimes all the $S$ curves are almost flat, as predicted by the narrow band Hubbard model, for semiconductors and metals, at sufficiently high temperatures [26,27]. This trend has been observed also in single crystals [28] and indicates that in iron chalcogenides transport has a more localized character than in other Fe-based families. The negative $S$ value is in sharp contrast with the corresponding positive value of $R_H$ in the same samples; again the multiband character of transport is crucial in accounting for this sign anomaly. The Hall resistance is dominated by the sign of $(\mu_h^2 n_h - \mu_e^2 n_e)$, whereas the Seebeck effect is dominated by the sign of $n_h\mu_h|S_h|-n_e\mu_e|S_e|$ according to eq. (1) and (3) respectively. Assuming that $\mu_h<\mu_e$, and $N_h(0)\approx N_e(0)$ as suggested by ab initio calculations [29], from the leading term of eq. (2) we have $n_h\mu_h|S_h|-n_e\mu_e|S_e| \propto \mu_h N_h(0) - \mu_e N_e(0) <0$, and thus a negative Seebeck coefficient. At the same time $R_H$ turns out to be positive if $n_h$ is sufficiently larger than $n_e$, so that $\mu_h^2 n_h > \mu_e^2 n_e$.

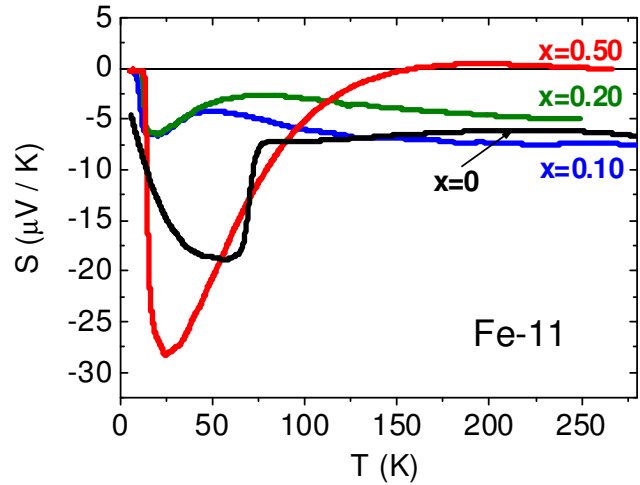

**Figure 7**. Seebeck coefficient versus T of selected samples of the Fe-11 series.

At $T_{SM}$ the Seebeck coefficient measured in the Fe$_{1+y}$Te sample undergoes an abrupt jump: consistently with Hall resistance data, a band rearrangement consequent to the structural transition occurs and the electron bands start to dominate the low temperature transport. Also in Se substituted samples the electron bands become more and more important with respect to the hole bands with decreasing temperature. However, the deep minimum of S above the onset of the superconducting transition is likely related to an excitation-drag mechanism. Indeed, it becomes more and more pronounced with increasing $T_c$, suggesting that excitations originating the superconducting pairing may also determines this drag term [22].

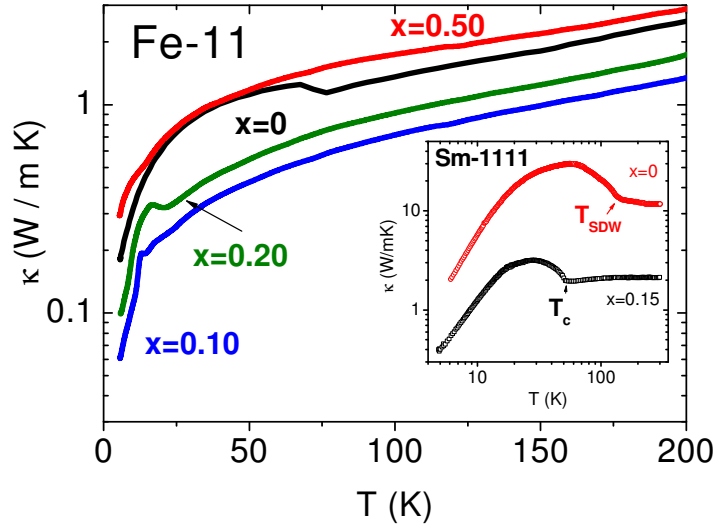

**Figure 8** : κ versus T in the Fe-11. Inset: κ versus T in Sm-1111.

A different probe of conduction mechanisms is provided by thermal conductivity measurements. The main panel of figure 8 shows κ in the Fe-11 whereas the inset reports data of Sm-1111. It can be easily shown that in these compounds thermal conductivity is dominated by phonons, being the electron contribution evaluated by the Wiedeman-Franz law negligible [17]. The main scattering mechanisms for phonons are carriers and structural defects, while intrinsic phonon-phonon scattering is dominant only in clean materials. Smaller κ values in Fe-11 than in Sm-1111 can be attributed to crystallographic disorder in the former, due to iron excess. Noteworthy the x=0.5 sample, with the least excess Fe within the series, exhibits the largest κ, Fe excess ions are mainly interstitial that should induce a higher degree of disorder than the one belonged to the Se-Te substitution. The FeTe sample has the next higher κ within the series displayed in figure 8; in this sample no substitutional disorder related to Se is present. The two samples with the largest κ, x=0 and x=0.5, are the ones where positive cyclotronic magnetoresistance is observed, confirming their lower degree of disorder.

A further feature, common to Fe-11 and Sm-1111 samples, emerges in figure 8: just below the ordering temperatures, either magnetic, structural and/or superconducting, the thermal conductivity sharply decreases. This means that below this temperature a scattering mechanisms for phonons is suppressed. In the Sm-1111 series, a detailed analysis of thermal data shows that the condensation of carries due to the opening of superconducting or SDW gaps, can account for the κ temperature behaviour [17].

In the Fe-11 series, a similar explanation could explain the κ increases observed in the doped samples below the respective superconducting transitions. On the contrary, the abrupt rise of κ below $T_{SM}$ for the FeTe sample can be hardly explained by the opening of a gap, because many investigations [8,9,10], suggest that the SDW in this compound has a gapless behaviour. In particular, Hall effect data in figure 5 support this view: no carrier condensation occurs at $T_{SM}$, but rather a band rearrangement such that hole transport dominates above $T_{SM}$, while electron transport dominates below it. Such crossover of carrier type is also evident in the Seebeck curve of the FeTe sample shown in figure 7. We argue that the hole band is the more coupled with phonons, so that at $T_{SM}$ the phonon scattering abruptly decreases as the hole band is depleted and the electron band is filled.

From the above transport data, it appears that in the Fe-11 series, multi band effects play a major role, and cannot be neglected in a quantitative analysis. The Fe(Te,Se) compounds have a more compensated and localized character than the $SmFeAs(O_{1-x}F_x)$ compunds.

## 5. Superconducting properties of Sm-1111 and Fe-11

The analysis of the transport data points out that the Sm-1111 samples, despite certainly multiband systems, can be analyzed in a single band framework to extract at least rough estimates of such parameters as carrier density and mobility from Hall effect and magnetoresistivity data. Only the analysis of Seebeck effect and in particular its behaviour as a function of doping requires the multiband character to be taken into account. Oppositely, in the case of the Fe-11 series, all transport properties such as Hall resistance, magnetoresistivity, Seebeck effect and thermal conductivity require a multiband approach to be understood, at least qualitatively. This points to a more compensated character of the Fe-11 system as compared to the Sm-1111 system. Another difference that emerges is the more localized transport in the Fe-11 system, likely related to the excess Fe that on one hand dopes electrons into the system and on the other hand introduces magnetic scattering centres. Scattering by localized magnetic impurities is clearly responsible for the *ln(T)* temperature dependence and for the negative magnetoresistance, both explained within a Kondo model. Such excess Fe changes in the different samples depending on the Se content, and in particular it decreases with increasing Se, which explains why isovalent Se substitution has the effect of doping holes into the system. Thus, Se substitution allows to tune hole and electron band contributions to all electrical and thermal transport properties.

We now try to identify how these results have a drawback on superconducting properties of these two families of superconductors.

From the above discussion it emerges the importance of the role played by excess Fe in Fe-11 which has a counterpart nor in Sm-1111 neither in other pnictides families. Magnetic moments of excess Fe should in principle produce pairbreaking, reducing $T_c$. In ref. [19] it has been observed that lowering the excess Fe concomitantly to doping with Se favors superconductivity, however no systematic investigation has been carried out on the specific role of excess iron on the superconducting properties. Here we try to correlate the huge upper critical field of the Fe-11 with magnetic scattering.

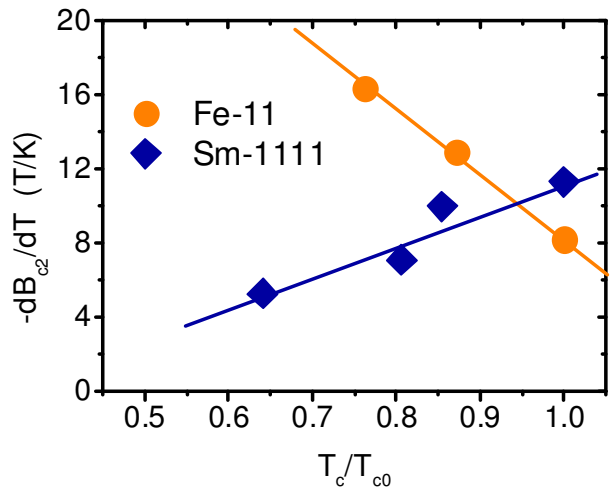

**Figure 9**: -$dB_{c2}/dT$ for the S-1111[13] and Fe-11 as a function of $T_c/T_{c0}$ where $T_{c0}$ is the critical temperature of the optimally doped sample ($T_{c0}$=51.5 K for Sm-1111 and $T_{c0}$=15.6 for Fe-11).

The slope of the upper critical field close $T_c$, $dB_{c2}/dT$, evaluated by magnetoresistivity measurements up to 9T are reported in table 1 for Fe-11. The values are rather large, comparable with those found in single crystals [30,11], taking into account that in polycrystalline samples the $B_{c2}$ parallel to the ab-planes is actually probed.

In figure 9 we compare $-dB_{c2}/dT$ for S-1111 [13] and Fe-11 as a function of $T_c/T_{c0}$ where $T_{c0}$ is the critical temperature of the optimally doped sample ($T_{c0}$=51.5 K for Sm-1111 and $T_{c0}$=15.6 for Fe-11). $-dB_{c2}/dT$ exhibits opposite behaviours for the two series: it increases with $T_c$ in Sm-1111 whereas it decreases in Fe-11.

The $B_{c2}$ slope within a Bardeen-Cooper-Shrieffer (BCS) approximation can be expressed by the following equation which describes the crossover between the clean and dirty limits:

$$\left|\frac{dB_{c2}}{dT}\right|_{T_c} \propto T_c\left(1+\frac{\xi_0}{l}\right) \quad (5)$$

where $\xi_0$ is the BCS coherence length, and $l$ is the carrier mean free path. Eq.(5) shows that in the clean limit ($\xi_0 < l$) a proportionality with $T_c$ is expected and this describes the behaviour of Sm-1111. In ref. [30] also for Fe-11 samples a clean superconductor scenario has been proposed on the basis of the exceptionally long mean free paths evaluated by a Drude model within a single band approximation; yet, the strongly compensated nature of Fe-11 discussed above makes such approach questionable. On the other hand the large $B_{c2}$ values and their relationship with $T_c$ exhibited by Fe-11 lead us to consider these compounds as dirty: indeed in this regime $\xi_0 > l$ and $|dB_{c2}/dT| \propto T_c(\xi_0/l) \propto 1/l$ independent of $T_c$ and increasing with decreasing $l$. Looking at figure 9, samples with lower $T_c$ should present shorter mean free path. The source of scattering cannot be Se which progressively diminishes with decreasing $T_c$, but excess Fe, which oppositely increases. Thus our data indicate that excess Fe drives the system into the dirty limit. This is only a suggestion that needs to be further investigated; however, it is interesting to consider the implication of this result. Dirty limit implies that magnetic scattering is effective in suppressing the mean free path more than in suppressing the $T_c$; this is quite unexpected for scattering by magnetic impurities. Indeed, the suppression of $T_c$ by impurity scattering is described by the equation $\ln(T_{c0}/T_c) = \psi(1/2+g) - \psi(1/2)$ [31], and is quantified by the reduced scattering rate parameter $g = \hbar\Gamma/2\pi k_B T_{c0} \approx (T_c/T_{c0})\xi_0/l$. The establishment of dirty limit for Fe-11 samples in figure 8 for which $T_c/T_{c0}\approx 1$ implies $g > 1$, while for scattering by magnetic impurities, $T_c$ should be rapidly suppressed for $g$ values lower than 1 [32]. The resilience of iron based compounds to scattering with magnetic impurities have been recently claimed by Tarantini *et al.* [33] which have analyzed the suppression of superconductivity in Nd-1111 single crystal irradiated by α-particles. In irradiated Nd-1111, which exhibit logarithmic upturn of resistivity and negative magnetoresistance like Fe-11, the superconductivity survives up to unusually high concentrations of magnetic and nonmagnetic defects produced by irradiation.

These results need to reconsider the interplay between superconductivity and magnetic scattering. From this point of view Fe-11 superconductors with magnetic local moment in proximity of the Fe layers offer the interesting opportunity for experimental investigation of this issue.

**Acknowledgement**
This work is partially supported by Compagnia di S. Paolo.